\documentclass[twoside]{article}
\usepackage{fleqn,espcrc2}
\usepackage{graphics}

\newcommand{\AmS}{{\protect\the\textfont2
  A\kern-.1667em\lower.5ex\hbox{M}\kern-.125emS}}

\title{Potential analysis of ${\cal N}=2$ SUSY gauge theory 
with the Fayet-Iliopoulos term} 

\author{Masato Arai\address{Department of Physics, 
                            Tokyo Metropolitan University,
                            Hachioji, Tokyo 192-0397, Japan \\}
        \thanks{Talk given at the D.V. Volkov Memorial Conference 
                ``Supersymmetry and Quantum Field Theory'', July 25-30,
                2000, Kharkov.}
        and Nobuchika Okada\address{Theory Group, KEK, Tsukuba, 
                            Ibaraki 305-0801, Japan}
}

\begin{document}

\begin{abstract}
We analyze the vacuum structure of spontaneously broken 
${\cal N}=2$ supersymmetric gauge theory 
 with the Fayet-Iliopoulos term. 
Our theory is based on the gauge group $SU(2) \times U(1)$ 
 with $N_f=2$ massless quark hypermultiplets 
 having the same $U(1)$ charges. 
In the classical potential, there are degenerate vacua 
 even in the absence of supersymmetry. 
It is shown that this vacuum degeneracy is smoothed out, 
 once quantum corrections are taken into account. 
While there is the runaway direction in the effective potential, 
 we found the promising possibility 
 that there appears the local minimum 
 with broken supersymmetry at the degenerate dyon point.
\vspace{1pc}
\end{abstract}

\maketitle

\section{Introduction}

There has been much progress in understanding the dynamics
 of strongly coupled supersymmetric (SUSY) gauge theories. 
Seiberg and Witten derived the exact low energy Wilsonian effective action 
 for ${\cal N}=2$ SUSY $SU(2)$ Yang-Mills theory \cite{s-w1}, 
 and generalized their discussion to the case 
 with up to four massive quark hypermultiplets \cite{s-w2}. 
The key ingredients which allow us to derive the exact results
 are duality and holomorphy. 

The results by Seiberg and Witten were extended to the case 
 with the explicit soft SUSY breaking terms by using spurion technique. 
Unless these terms do not change the holomorphy and duality properties 
 of the theory, we can derive the exact effective action 
 for ${\cal N}=1$ and ${\cal N}=0$ (non-supersymmetric) SUSY gauge theories 
 up to the leading order for the soft SUSY breaking terms. 
In Refs. \cite{peskin}, the exact superpotential and 
 the phase structure in ${\cal N}=1$ SQCD were discussed 
 based on the ${\cal N}=2$ SUSY gauge theory 
 with some soft breaking terms. 
In Refs. \cite{marino1,marino2,marino3}, the vacuum structure 
 of non-SUSY gauge theory was investigated 
 in which soft SUSY breaking terms directly break 
 ${\cal N}=2$ SUSY to ${\cal N}$=0. 

In this paper, we study a spontaneously broken ${\cal N}=2$ 
 SUSY gauge theory. 
It is well known that, in the frame work of ${\cal N}=2$ 
 SUSY theory, the only possibility to break SUSY spontaneously 
 is to introduce the Fayet-Iliopoulos (FI) term. 
Therefore, in the following, we consider the gauge theory which includes
 $U(1)$ gauge interaction together with the FI term. 

The simplest example of this type of theory is 
 ${\cal N}=2$ SUSY QED (SQED) with the FI term \cite{arai}. 
At the classical level, although SUSY is spontaneously broken
 in Coulomb branch, 
 there are degenerate vacua (moduli space) 
 parameterized by the vacuum expectation value 
 of the scalar field, $a$, in the $U(1)$ vectormultiplet. 
The direction of this vacuum degeneracy in the absence of SUSY 
 is called ``pseudo flat'' direction. 
However, it is expected that this direction is lifted up, 
 once quantum corrections are taken into account. 
By virtue of ${\cal N}=2$ SUSY, the effective action is found to be 
 one loop exact, and the effective gauge coupling is given by 
 $e(a)^2\sim 1/\log (\Lambda_L/ a )$,
 where $\Lambda_L$ is the Landau pole. 
Note that there are two singular regions in moduli space, namely, 
 the ultraviolet (UV) region such as $|a| \geq \Lambda_L$ 
 and the massless singular point at the origin $a=0$. 
Since the effective potential is described as $V\sim e(a)^2$, 
 the potential minimum emerges at the origin, 
 where SUSY is formally restored. 
However, since this point is the singular point, 
 we conclude that there is no well-defined vacuum in this theory. 

In this paper, we investigate the vacuum structure 
 of more interesting theory with spontaneous ${\cal N}=2$ SUSY breaking.
\footnote{ 
For complete analysis including the $N_f=1$ case, 
 see our original paper \cite{arai2}.
}
Our theory is based on the gauge group $SU(2)\times U(1)$ 
 with $N_f=2$ massless quark hypermultiplets 
 having the same $U(1)$ charges.  
In the UV region, the behavior of the effective potential 
 can be well understood based on the perturbative discussion, 
 since the $SU(2)$ gauge interaction is weak there. 
On the other hand, it is expected that the behavior of the effective 
 potential in the infrared region is drastically changed compared 
 with SQED, because of the presence of the $SU(2)$ gauge dynamics.

\section{Classical structure of ${\cal N}=2$ $SU(2)\times U(1)$ gauge theory}

In this section, we briefly discuss the classical structure of our theory. 
The analysis of the classical potential was 
 originally addressed in Ref. \cite{fayet}. 

We describe the classical Lagrangian in terms of ${\cal N}=1$ 
 superfields: adjoint chiral superfield $A_i$ 
 and superfield strength $W_i$ 
 in the vectormultiplet ($i=1,2$ denote the $U(1)$ 
 and the $SU(2)$ gauge symmetries, respectively),  
 and two chiral superfields $Q^i_\alpha$ and $\tilde{Q}_i^\alpha$ 
 in the hypermultiplet ($i=1,2$ is the flavor index, 
 and $\alpha=1,2$ is the $SU(2)$ color index). 
The classical Lagrangian is given by 
\begin{eqnarray}
{\cal L}&=&{\cal L}_{HM}
           +{\cal L}_{VM}
           +{\cal L}_{FI} \; ,  \label{eq:lag} \\
{\cal L}_{HM}
   &=& \int d^4\theta  \left( 
       Q_i^\dagger e^{2V_2+2V_1} Q^i\right. \nonumber\\
   & & \left.~~~~~~~~~~~+\tilde{Q}_i e^{-2V_2-2V_1} \tilde{Q}^{\dagger i}
          \right)  \\  
   &+& \sqrt{2} \left( \int d^2 \theta 
           \tilde{Q}_i \left( A_2+A_1 \right) Q^i + h.c. \right), \\
{\cal L}_{VM} 
        &=&\frac{1}{2\pi} \mbox{Im}\left[ \mbox{tr}
           \left\{\tau_{22}
           \left( \int d^4\theta A_2^\dagger e^{2 V_2} A_2 e^{-2 V_2}
           \right.\right.\right. \nonumber \\
        & &\left.\left.\left.~~~~~~~~~~~~~~+\frac{1}{2}\int d^2 \theta  W_2^2 
           \right)\right\}\right] \nonumber \\
   &+& \frac{1}{4\pi}{\rm Im}\left[\tau_{11} \left(\int d^4\theta 
           A_1^\dagger A_1\right.\right. \; \\
        & &\left.\left.~~~~~~~~~~~~~~~+\frac{1}{2} \int d^2\theta 
           W_1^2 \right)\right] \; , \\
{\cal L}_{FI} 
        &=&\int d^4\theta \xi V_1  \; ,      \label{eq:FI}
\end{eqnarray}
where $\tau_{22}=i\frac{4\pi}{g^2}+\frac{\theta}{2\pi}$ 
 and $\tau_{11}=i\frac{4\pi}{e^2}$ are the gauge couplings 
 of the $SU(2)$ and the $U(1)$ gauge interactions, respectively.
Here we take the notation, 
 $T(R) \delta^{ab}$=tr($T^a T^b)=\frac{1}{2}\delta^{ab}$. 
The same $U(1)$ charges of the hypermultiplets are normalized to be one.
The last term in (\ref{eq:lag}) is the FI term
 with the coefficient $\xi$ of mass dimension two. 

From the above Lagrangian, the classical potential is read off as
\begin{eqnarray} 
 V &=&\frac{1}{g^2}\mbox{\rm tr}[A_2, A_2^\dagger]^2 
       +\frac{g^2}{2}  
 ( q_i^{\dagger} T^a q^i- \tilde{q}_i T^a  \tilde{q}^{\dagger i} )^2  
 \nonumber \\
    &+& q_i^\dagger [A_2, A_2^\dagger] q^i  
     - \tilde{q}_i [A_2,A_2^\dagger] \tilde{q}^{\dagger i} 
    +2 g^2 | \tilde{q}_i T^a q^i |^2  \nonumber \\
   &+&\frac{e^2}{2} \left( \xi+ q_i^\dagger q^i- \tilde{q}_i 
       \tilde{q}^{\dagger i} \right)^2 
     +2 e^2 |\tilde{q}_i q^i|^2 \nonumber \\
   &+& 2 \left( q_i^\dagger |A_2+A_1|^2 q^i 
    +\tilde{q}_i |A_2+A_1|^2 \tilde{q}^{\dagger i} \right),
\end{eqnarray}
where $A_2$, $A_1$, $q^i$ and $\tilde{q}_i$ are scalar components 
 of the corresponding chiral superfields, respectively. 
The potential minimum is obtained by solving 
 the stationary conditions with respect to these scalar components. 
There are some solutions, and one example is given by
\begin{eqnarray}
|q^i|^2 = 0, \; \; 
     |\tilde{q}_1|^2=  \frac{e^2}{\frac{1}{4}g^2+e^2} \xi, \; \; 
     |\tilde{q}_j|^2= 0 \; \; (j \neq 1), \nonumber   
\end{eqnarray}
\begin{eqnarray}
A_2 + A_1 &=& \left(\begin{array}{cc}
              \frac{a_2}{2} & 0 \\  0 & -\frac{a_2}{2}  
                   \end{array}\right)
              +\left(\begin{array}{cc}
                     a_1 & 0 \\  0 & a_1  
                     \end{array}\right)
               \nonumber \\
          &=& \left(\begin{array}{cc} 0 & 0 \\
                     0 & z  \end{array}\right),
        \label{eq:mix}
\end{eqnarray}
where $a_1$ and $a_2$ are complex parameters, 
 and $z$ is arbitrary constant.  
In this example, the gauge symmetry $SU(2)\times U(1)$ is broken to $U(1)$. 
The potential energy is given by 
\begin{eqnarray}
V=\frac{\xi^2}{2}
           \frac{e^2 g^2}{4e^2+g^2}.
\end{eqnarray}

Note that the classical potential has the pseudo flat direction  
 parameterized by $a_1$ or $a_2$ 
 with the condition $a_1+\frac{1}{2}a_2=0$. 
We expect that this direction is lifted up, 
 once quantum corrections are take into account, 
 and the true non-degenerate vacuum is selected out 
 after the effective potential is analyzed. 
This naive expectation seems natural, 
 if we notice that the above potential energy is 
 described by the bare gauge couplings, 
 which should be replaced by the effective one 
 (non-trivial functions of moduli parameters) 
 in the effective theory.

\section{Quantum structure of ${\cal N}=2~SU(2)\times U(1)$ gauge theory}

\subsection{Effective Action}

In this subsection, we describe the low energy Wilsonian effective 
 Lagrangian of our theory. 
If we could completely integrate the action to zero momentum, 
 the exact effective Lagrangian ${\cal L}_{EXACT}$ could be obtained, 
 which is described by light fields, the dynamical scale 
 and the coefficient of the FI term $\xi$. 
However, this is highly non-trivial and very difficult task. 
In the following discussion, suppose that the coefficient $\xi$, 
 the order parameter of SUSY breaking, 
 is much smaller than the dynamical scale of the $SU(2)$ gauge interaction. 
Then we consider the effective action up to the leading order of $\xi$. 
The exact effective Lagrangian, if it could be obtained, 
 can be expanded with respect to the parameter $\xi$ as 
\begin{eqnarray}
{\cal L}_{EXACT}
        ={\cal L}_{SUSY}
         +\xi {\cal L}_1 + {\cal O}(\xi^2).
         \label{eq:eff}
\end{eqnarray}
Here, the first term ${\cal L}_{SUSY}$ 
 is the exact effective Lagrangian containing full SUSY quantum corrections. 
The second term is the leading term of $\xi$, 
 and nothing but the FI term at tree level.
Analyzing the effective Lagrangian up to the leading order of $\xi$,
 we obtain the effective potential of the order of $\xi^2$. 
The coefficient of $\xi^2$ in the effective potential includes 
 full SUSY quantum corrections. 
Therefore, what we need to analyze the effective potential 
 is nothing but the effective Lagrangian 
 ${\cal L}_{SUSY}$. 

Except the FI term, the classical $SU(2)\times U(1)$ gauge theory 
 has moduli space, which is parameterized by $a_2$ and $a_1$. 
On this moduli space except the origin, the gauge symmetry is broken to 
 $U(1)_c\times U(1)$. 
Here $U(1)_c$ denotes the gauge symmetry in the Coulomb phase 
 originated from the $SU(2)$ gauge symmetry. 
Before discussing the effective action of this theory, 
 we should make it clear how to treat the $U(1)$ gauge interaction part. 
In the following analysis, this part is, as usual, discussed 
 as a cut-off theory.
Thus, the Landau pole $\Lambda_L$ is inevitably introduced 
 in our effective theory, and the defined region 
 of the moduli parameter $a_1$ is constrained 
 within the region $|a_1|< \Lambda_L$. 
According to this constraint, the defined region for moduli parameter 
 $a_2$ is found to be also constrained in the same region, 
 since two moduli parameters are related with each other 
 through the hypermultiplets. 
We take the scale of $\Lambda_L$ to be much larger than 
 the dynamical scale of the $SU(2)$ gauge interaction $\Lambda_2$, 
 so that the $U(1)$ gauge interaction is always weak 
 in the defined region of moduli space. 
Note that, in our framework, we implicitly assume that 
 the $U(1)$ gauge interaction has no effect on the $SU(2)$ gauge dynamics.
This assumption is justified in the following discussion about 
 the monodromy transformation (see Eq.~(\ref{eq:mono})). 

We first discuss the general formulae for the effective Lagrangian 
 ${\cal L}_{SUSY}$, which 
 consists of two parts described by light vectormultiplets 
 and hypermultiplets, 
 ${\cal L}_{SUSY}={\cal L}_{VM}+{\cal L}_{HM}$. 
The vectormultiplet part ${\cal L}_{VM}$, 
 which is consistent with ${\cal N}=2$ SUSY and all the symmetries 
 in our theory, is given by  
\begin{eqnarray}
{\cal L}_{VM}
        &=&\frac{1}{4\pi}
           \mbox{\rm Im}\Bigg{\{}
           \sum_{i,j=1}^{2}\left(\int d^4\theta
           \frac{\partial F}{\partial A_i} A_i^\dagger 
           \right.\nonumber \\
        & &~~~~~~~~~~~\left.+ \int d^2\theta \frac{1}{2} 
                  \tau_{ij} W_i W_j \right) \Bigg{\}} ,
\end{eqnarray}
where $F(A_2,A_1,\Lambda_2,\Lambda_L)$ 
 is the prepotential, which is the function of moduli parameters  
 $a_2$, $a_1$, the dynamical scale $\Lambda_2$, 
 and the Landau pole $\Lambda_L$. 
The effective gauge coupling $\tau_{ij}$ is defined as 
\begin{eqnarray}
\tau_{ij}=
   \frac{\partial^2 F}{\partial a_i \partial a_j} \; ~~(i=1,2),   
   \label{eq:coupling}
\end{eqnarray}
The part ${\cal L}_{HM}$ is 
 described by a light hypermultiplet 
 with appropriate quantum number $(n_m,n_e)_n$, 
 where $n_m$ is magnetic charge, $n_e$ is electric charge, 
 and $n$ is the $U(1)$ charge.  
This part should be added to the effective Lagrangian 
 around a singular point on moduli space, 
 since the hypermultiplet is expected to be light there
 and enjoys the correct degrees of freedom in the effective theory. 
The explicit description is given by 
\begin{eqnarray}
{\cal L}_{HM}
   &=& \int d^4 \theta \left( 
      M^\dagger e^{2 n_m V_{2D} + 2 n_e V_2+ 2 n V_1}M
      \right.\nonumber \\
   &+&\left.\tilde{M} e^{-2 n_m V_{2D} -2 n_e V_{2} - 2 n V_1}
      \tilde{M}^{\dagger} \right) \nonumber \\
   &+&\sqrt{2}\Big{(} \int d^2\theta \tilde{M} 
      (n_m A_{2D} +n_e A_2 \nonumber \\
   & &~~~~~~~~~~~~~~~~~~~+ n A_1) M+h.c. \Big{)}, 
\end{eqnarray}
where $M$ and $\tilde{M}$ denote light quark or light dyon 
 hypermultiplet~(the BPS states), and $V_{2D}$ is the dual 
 gauge field of $U(1)_c$. 

In order to obtain an explicit description of the effective Lagrangian, 
 let us consider the monodromy transformation of our theory. 
Suppose that moduli space is parameterized by the vectormultiplet scalars 
 $a_2$, $a_1$ and their duals $a_{2D}$, $a_{1D}$ 
 which are defined as $a_{iD}=\partial F/ \partial a_i$ ($i=1,2$). 
These variables are transformed into their linear combinations 
 by the monodromy transformation. 
In our case, the monodromy transformation is subgroup 
 of $Sp(4,\mbox{\bf R})$, which leaves the effective Lagrangian invariant, 
 and the general formula is found to be \cite{marino3} 
\begin{eqnarray}
 \left(\begin{array}{c}
       a_{2D} \\  a_{2}  \\  a_{1D} \\  a_1  \end{array}  \right)
 \rightarrow 
 \left(\begin{array}{c}
       \alpha a_{2D}+\beta a_2+p a_1   \\
       \gamma a_{2D}+\delta a_2+q a_1  \\
       a_{1D}+pa_{2p}-qa_{2q}-pqa_1\\ a_1 
\end{array}  \right),  
  \label{eq:mono}
\end{eqnarray}
where $a_{2p}=\gamma a_{2D}+\delta a_2,
~a_{2q}=\alpha a_{2D}+\beta a_2$, 
$
\left(\begin{array}{cc}
\alpha & \beta  \\
\gamma & \delta 
      \end{array}
\right)
\in
SL(2,\mbox{\bf Z})
$,
 and $p,q \in\mbox{\bf Q}$. 
Note that this monodromy transformation for the combination 
 $(a_{2D},a_{2},a_1)$ is exactly the same as that 
 for SQCD with massive quark hypermultiplets, 
 if we regard $a_1$ as the same mass of the hypermultiplets 
 such that $m= \sqrt{2} a_1$.  
This means that the $U(1)$ gauge interaction part plays the only role 
 as the mass term for the $SU(2)$ gauge dynamics. 
This observation is consistent with our assumption. 
On the other hand, the $SU(2)$ dynamics plays an important role 
 for the $U(1)$ gauge interaction part, 
 as can be seen in the transformation law of $a_{1D}$. 
This monodromy transformation is also used to derive dual variables 
 associated with the BPS states. 
As a result, the prepotential of our theory turns out to be 
 essentially the same as the result in \cite{s-w2} 
 with understanding the relation $A_1 = m/\sqrt{2}$, 
\begin{eqnarray}
& &F(A_2,A_1,\Lambda_{2},\Lambda_L)\nonumber \\
& &~~=F_{SU(2)}^{(SW)}(A_2,m,\Lambda_{2})
    \Big{|}_{A_1=\frac{m}{\sqrt{2}}}+C A_1^2,
    \label{eq:pre}
\end{eqnarray}
where the first term is the prepotential of ${\cal N}=2$ SQCD 
 with hypermultiplets having the same mass $m$, 
 and $C$ is free parameter. 
The freedom of the parameter $C$ is used to determine 
 the scale of the Landau pole 
 relative to the scale of the $SU(2)$ dynamics.  
\begin{figure*}
\hspace{20mm}
\scalebox{0.6}[0.6]
{\includegraphics{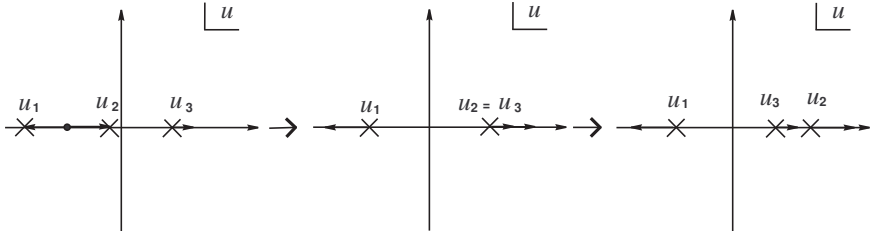}}
\caption{The flow of the sigular points in $N_f=2$ case with 
the same mass.}
\label{flow}
\end{figure*}

\subsection{Effective Potential}

The effective potential can be read off from the above Lagrangian 
 with the FI term. 
Eliminating the auxiliary fields by using the equations of motion, 
 we obtain
\begin{eqnarray}
V &=& \frac{b_{22}}{2\det b}\xi^2
      +S(a_2,a_1)\left\{(|M|^2 - |\tilde{M}|^2)^2
      \right.\nonumber \\
  &+& \left.4|M \tilde{M}|^2\right\}
      +2 T(a_2,a_1)(|M|^2 + |\tilde{M}|^2) 
      \nonumber \\
  &-& U(a_2,a_1) (|M|^2-|\tilde{M}|^2),
\end{eqnarray}
where $S$, $T$ and $U$ are defined as 
\begin{eqnarray}
S(a_2,a_1)&=&\frac{1}{2b_{22}}
            +\frac{(b_{12}-nb_{22})^2}
                  {2b_{22}\det b}, \\
T(a_2,a_1)&=&|n_ma_{2D}+n_ea_2+na_1|^2, \\
U(a_2,a_1)&=&\frac{b_{12}-nb_{22}}{\det b}\xi.
\end{eqnarray}
Solving the stationary conditions with respect to the 
hypermultiplet, we obtain three solutions as follows: 
\begin{eqnarray}
1.~& & M = \tilde{M} =0;  
       ~~~~V=\frac{b_{22}}{2\det b}\xi^2,  \label{eq:sol1} \\
2.~& & |M|^2=-\frac{2T-U}{2S}, \; \tilde{M}=0; \nonumber \\
   & & ~~~~~~~V=\frac{b_{22}}{2\det b}\xi^2-S |M|^4,
       \label{eq:sol2} \\
3.~& & M=0, \; |\tilde{M}|^2 = -\frac{2T+U}{2S};\nonumber \\
   & & ~~~~~~~V=\frac{b_{22}}{2\det b}\xi^2 
          -S |\tilde{M}|^4 .  \label{eq:sol3}
\end{eqnarray}
The solution (\ref{eq:sol2}) or (\ref{eq:sol3}), 
 in which the light hypermultiplet acquires 
 the vacuum expectation value, 
 is energetically favored, because of $\det b >0$ and $S(a_2,a_1)>0$.  
Since the hypermultiplet appears in the theory as the light BPS state  
 around the singular point on moduli space, 
 the potential minimum is expected to emerge there. 
On the other hand, the solution (\ref{eq:sol1}) describes 
 the potential energy away from the singular points, 
 which smoothly connects to the solution (\ref{eq:sol2}) or (\ref{eq:sol3}). 

\subsection{Periods and Effective Couplings}

It was shown that the effective potential is described by the periods 
 $a_{2D}$, $a_{2}$ and the effective coupling $b_{ij}$. 
In this subsection, we derive the periods and the effective couplings 
 in order to give an explicit description of the effective potential. 

As already discussed, the periods $a_{2D}$ and $a_2$ 
 are the same as that of SQCD.
The periods are defined as the contour integrals  
$
a_{2D}=\oint_{\alpha_1}\lambda,
~~a_{2}=\oint_{\alpha_2}\lambda,
$
where $\lambda$ is a meromorphic differential on the algebraic curve, 
 and the cycles $\alpha_1$ and $\alpha_2$ are defined 
 so as to encircle the roots of the algebraic curve $e_2$ and $e_3$, 
 and $e_1$ and $e_3$, respectively.
In our case, the roots are given by
\begin{eqnarray}
e_1&=&\frac{u}{24}-\frac{\Lambda_2^2}{64}\nonumber \\
   & &~~~-\frac{1}{8}\sqrt{u+\frac{\Lambda_2^2}{8}
                  +\Lambda_2 m}
                 \sqrt{u+\frac{\Lambda_2^2}{8}
                  -\Lambda_2 m},
                 \nonumber \\
e_2&=&\frac{u}{24}-\frac{\Lambda_2^2}{64}\nonumber \\
   & &~~~+\frac{1}{8}\sqrt{u+\frac{\Lambda_2^2}{8}
                  +\Lambda_2 m}
                 \sqrt{u+\frac{\Lambda_2^2}{8}
                  -\Lambda_2 m},
                 \nonumber \\
e_3&=&-\frac{u}{12}+\frac{\Lambda_2^2}{32}. 
                 \label{eq:root2}
\end{eqnarray}
Then, the periods are described as \cite{ferrari}
($a_{2D}$ and $a_2$ are denoted
 by $a_{21}$ and $a_{22}$, respectively)
\begin{eqnarray}
& & a_{2i}=-\frac{\sqrt{2}}{4\pi}
    \left(-\frac{4}{3}uI_1^{(i)}+8I_2^{(i)}\right.
    \nonumber \\
& &~~~\left.+\frac{m^2\Lambda_2^2}{8}
    I_3^{(i)}
    \left(-\frac{u}{12}
    -\frac{\Lambda_2^2}{32}\right)\right)
    -\frac{m}{\sqrt{2}}\delta_{i2},
    \label{eq:period2} 
\end{eqnarray}
with the integral $I_i^{(1)}\; (i=1,2,3)$  given by
\begin{eqnarray}
I_1^{(1)}&=&\int_{e_2}^{e_3}\frac{dX}{Y}
          = \frac{iK(k^\prime)}{\sqrt{e_2-e_1}}, 
            \label{eq:formula1} \\
I_2^{(1)}&=&\int_{e_2}^{e_3}\frac{XdX}{Y}
          = \frac{ie_1}{\sqrt{e_2-e_1}}K(k^\prime)
             \nonumber \\
         & &~~~~~~~~~~+i\sqrt{e_2-e_1}E(k^\prime), 
             \label{eq:formula2} \\
I_3^{(1)}&=&\int_{e_2}^{e_3}\frac{dX}{Y(X-c)}
             \nonumber \\
         &=& \frac{-i}{(e_2-e_1)^{3/2}}
            \left\{\frac{K(k^\prime)}{k+\tilde{c}}\right.
             \nonumber \\
         & &~~\left.+\frac{4k}{1+k}
            \frac{1}{\tilde{c}^2-k^{2}}
            \Pi_1\left(\nu,\frac{1-k}{1+k}         
                \right)            
            \right\},
            \label{eq:formula3}
\end{eqnarray}
where $k^2 = \frac{e_3-e_1}{e_2-e_1}$, 
      $k^{\prime 2}=1-k^2=\frac{e_2-e_3}{e_2-e_1}$,  
      $c=-\frac{u}{12}-\frac{\Lambda_2}{32}$ 
      is the pole of the meromorphic differential, 
      $\tilde{c}= \frac{c-e_1}{e_2-e_1}$, 
  and $\nu=-\left(\frac{k+\tilde{c}}{k-\tilde{c}}\right)^2
          \left(\frac{1-k}{1+k}\right)^2$. 
The formulae for $I_i^{(2)}$ are obtained from $I_i^{(1)}$ 
 by exchanging the roots, $e_1$ and $e_2$. 
In Eqs.~(\ref{eq:formula1})-(\ref{eq:formula3}), 
 $K$, $E$, and $\Pi_1$ are the complete elliptic integrals \cite{higher}. 
\begin{figure*}
\hspace{20mm}
\scalebox{0.6}[0.6]
{\includegraphics{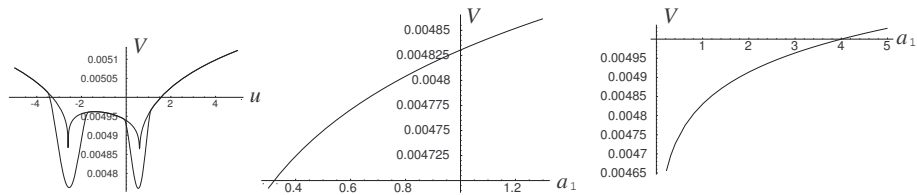}}
\caption{The left figure shows the potential for $a_1=0.4$.
The middle and right figures show the evolutions of the potential 
energies at the singular points $u_2$ and $u_3$, respectively.}
\label{fix}
\end{figure*}

Next we give the effective couplings defined 
 as (\ref{eq:coupling}).
After some calculations, we obtain 
\begin{eqnarray}
\tau_{22}&=&\frac{\omega_1}{\omega_2},\\
\tau_{12}&=&-\frac{2z_0}{\omega_2},
         \label{eq:tau12} \\
\tau_{11}&=&-\frac{1}{\pi i}
            \left[\log\sigma(2z_0)
           +\frac{4z_0^2}{\omega_2}
           I_2^{(2)}\right] + C ,  
         \label{eq:tau11}
\end{eqnarray}
where $\omega_i$ is the period of the Abelian differential, 
 $\omega_i=\oint_{\alpha_i}\frac{dX}{Y}=2I_1^{(i)}~(i=1,2)$, 
 $z_0=-\frac{1}{\sqrt{e_2-e_1}}F(\phi,k)$
 ($F(\phi,k)$ is the incomplete elliptic integral of 
 the first kind with $\sin^2\phi=\frac{e_2-e_1}{c-e_1}$), 
 $\sigma$ is the Weierstrass sigma function, 
 and $C$ is the constant in Eq.~(\ref{eq:pre}). 

Note that, since the gauge coupling $b_{11}$ is found to be 
 a monotonically decreasing function of large $|a_1|$ 
 with fixed $u$, and vice versa,  
 the scale of the Landau pole is defined as $|a_1|=\Lambda_L$ 
 at which $b_{11}=0$. 
The large $\Lambda_L$ required by our assumption 
 is realized by taking appropriate value.
In the following analysis, we fix $C = 4\pi i$, 
 which corresponds to $\Lambda_L \sim 10^{18}$ 
 for fixed $\Lambda_{2}=2\sqrt{2}$ 

\section{Potential Analysis}

Based on the results given by the previous sections, 
 now let us investigate the vacuum structure of our theory. 
Since the effective potential is the function of 
 two complex moduli parameters $u$ and $a_1$, 
 it is a very complicated problem to figure out 
 behaviors of the effective potential 
 in the whole parameter space. 
However, note that, for our aim, it is enough 
 to evaluate the potential energy just around each singular points, 
 because these points are energetically favored 
 (see Eqs.~(\ref{eq:sol1})-(\ref{eq:sol3})). 
The singular points on the moduli space parameterized by $u$ 
 flow according to the variation of $a_1$. 
In the following discussion, we evaluate the effective potential 
 along the flow of the singular points, 
 and examine which point is energetically favored 
 on the line of the flow.  
\begin{figure*}
\hspace{20mm}
\scalebox{0.6}[0.6]
{\includegraphics{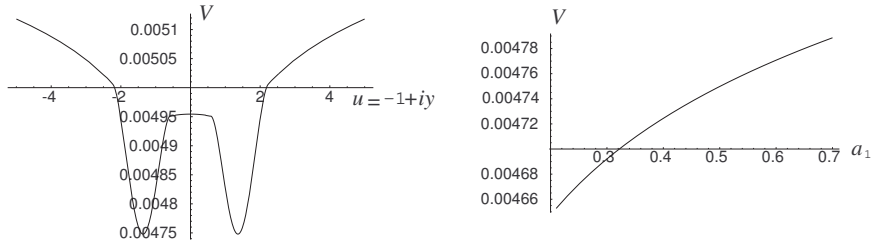}}
\caption{The effective potential~(left)
for $a_1=i\frac{\sqrt{2}}{4}, \xi=0.1$ 
along $u=-1$ and the evolution of the 
minimum~(right).}
\label{flavor2-4}
\end{figure*}

Let us first consider the flow of the singular points.  
The discriminant of the algebraic curve 
 can be easily solved and leads to the three singular points 
 such that 
 $u_1=-\sqrt{2}a_1\Lambda_2-\frac{\Lambda_2^2}{8}$,
 $ u_2=\sqrt{2}a_1\Lambda_2-\frac{\Lambda_2^2}{8}$ 
 and $ u_3=(\sqrt{2}a_1)^2+\frac{\Lambda_2^2}{8}$. 
We investigate the case $\mbox{Im} a_1 =0$, for simplicity.  
The flow of the singular points is sketched in Fig.~\ref{flow}. 
For $a_1=0$, the singular points appear 
 at $u_1=u_2=-1$ and $u_3=1$. 
Here, at $u=-1$, two singular points degenerate. 
For non-zero $a_1>0 $, 
\footnote{ We consider only the case $a_1 > 0$, 
 since the result for $a_1 < 0$ can be obtained 
 by exchanging $u_1 \leftrightarrow u_2$,
 as be seen from the solution of the discriminant.} 
 this singular point splits into two singular points $u_1$ and $u_2$, 
 which corresponds to the BPS states with quantum numbers 
 $(1,1)_{-1}$ and $(1,1)_1$, respectively.
As $a_1$ is increasing, these singular points, $u_1$ and $u_2$, 
 are moving to the left and the right on real $u$-axis, respectively.  
Two singular points, $u_2$ and $u_3$, collide 
 at $u=3\Lambda_2^2/8$ for $a_1= \Lambda_2/ (2 \sqrt{2})$.
This collision point is called Argyres-Douglas (AD) point \cite{douglas}, 
 at which the theory is believed to transform 
 into ${\cal N}=2$ superconformal theory. 
As $a_1$ is increasing further, there appear two singular points 
 $u_2$ and $u_3$ again, 
 and quantum numbers of the corresponding BPS states, 
 $(1,1)_1$ and $(0,1)_0$, change into $(1,0)_1$ and $(1,-1)_1$, 
 respectively. 
The singular point $u_2$ is moving to the right faster than $u_3$. 

We investigate the vacuum structure by varying the values of $a_1$. 
For $ a_1=0.4$, the effective potential is plotted in 
 Fig.~\ref{fix}~(left). 
While there appear the potential minima 
 at two singular points $u_1$ and $u_2$, 
 the monopole condensation is too small for the potential 
 to have a minimum at the singular point $u_3$. 
The top figure with the cusps shows the effective potential 
 without the dyon condensations, 
 and the bottom figure shows one with the condensations. 
Note that the cusps are smoothed out in the bottom figure. 
This means that the dyons enjoy the correct degrees of freedom 
 in our effective theory around the singular point. 
The evolutions of the values of the potential minimum 
 for the singular points $u_2$ and $u_3$ are depicted 
 in Fig.~\ref{fix} (middle and right).  
We find that both of two minima go down to the point $a_1=0$, 
 and thus the effective potential is bounded from below,  
 at least, along real $u$-axis. 

Next, we examine whether the effective potential is bounded 
 in all the directions for general complex $a_1$ values. 
As an example, let us consider the case $\mbox{Re}a_1=0$. 
For $a_1 \neq 0$, the two singular points $u_1$ and $u_2$  
 appear on the imaginary $u$-axis with $\mbox{Re} u =-1$. 
The effective potential is depicted in Fig.~\ref{flavor2-4} 
 along this axis for $a_1= i \frac{\sqrt{2}}{4}$. 
There appear two potential minima on the singular points. 
The right figure shows the evolution of the values 
of one potential minimum,
\footnote{ Two potential minima are degenerate, 
 since the effective potential has the CP symmetry 
 under the exchange $u \leftrightarrow u^\dagger$.}
and we find that the effective potential 
 is also bounded in this 
 direction. 
We can check that the effective potential is bounded from below 
 for all the values of small $|a_1|$. 
Therefore, there is a possibility that the effective potential has 
 the local minimum at the points $u=-1$ and $a_1=0$.   

\begin{figure}[h]
\hspace{5mm}
\scalebox{0.65}[0.65]
{\includegraphics{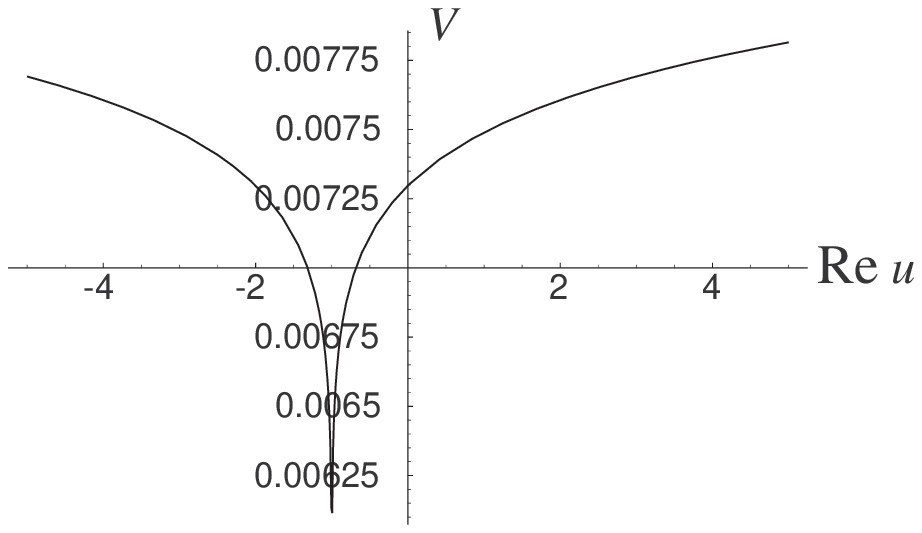}}
\caption{The effective potential without the 
contribution of the dyon condensations in the limit 
$a_1\rightarrow 0$.}
\label{flavor2-5}
\end{figure}

However, note that our description is not applicable for small $|a_1|$, 
 since the condensations 
 of two dyon states 
 are going to overlap with each other.
Unfortunately, we have no knowledge about the correct description 
 of the effective theory in this situation. 
Nevertheless, we conclude that there must appear the local minimum 
 with broken SUSY in the limit $a_1 \rightarrow 0$ 
 from the result in the following. 
For the limit $a_1 \rightarrow 0$, 
 the effective potential without the dyon condensations 
 is depicted in Fig.~\ref{flavor2-5}.
We can find that there appears the minimum at $u=-1$, 
 and the value of the effective potential on the cusp 
 is non-zero, $V \sim 0.0061$. 
If we had the correct description of the effective theory for $a_1$=0, 
 this cusp might be smoothed out.  
However, there is no reason that SUSY is restored at $u=-1$, 
 because the correct effective theory must have no singularity 
 for the Kahler metric.  
Therefore, there is the promising possibility of the appearance 
 of the local minimum with broken SUSY at $u=1$ and $a_1=0$.  

Finally, let us get back to discussion of the case $\mbox{Im} a_1=0$. 
The effective potential for $a_1 >  \frac{\Lambda_2}{2 \sqrt{2}}$ 
 has two minima only at two singular points $u_1$ and $u_2$. 
The monopole condensation is too small for the effective potential 
 to have a minimum at $u_3$. 
While the evolution of the value of the potential minimum 
 on $u_1$ is the same as for $0 < a_1 < \frac{\Lambda_2}{2 \sqrt{2}}$, 
 the value of the potential minimum on $u_2$ point 
 is monotonically decreasing, as $a_1$ is increasing. 
Thus, there is a runaway directions 
 along the flow of the quark singular point $u_2$.  
We can find that the runaway direction always 
 appears along the quark singular point 
 for general complex $a_1$ values.  

\section{Conclusion} 
We analyzed the vacuum structure of spontaneously broken 
 ${\cal N}=2$ SUSY gauge theory with the Fayet-Iliopoulos term. 
Our theory is based on the gauge group $SU(2) \times U(1)$ 
 with $N_f=2$ massless quark hypermultiplets 
 having the same $U(1)$ charges. 

We formulated the effective action up to the leading order 
 of the SUSY breaking parameter. 
Then the effective potential is obtained as the function of 
 the moduli parameters. 
Examining the minimum of the effective potential, 
 we found that the singular points are energetically favored, 
 because of the condensations of the light BPS 
 states. 
The singular points flow 
 according to the values of the moduli parameters. 
Thus, we analyzed the effective potential 
 along the flows of the singular points, 
 and examined which point is energetically favored 
 on the line of the flow.  
 
While there is the runaway directions 
 along the flow of the quark singular point, 
 we found the promising possibility that 
 the local minimum with broken SUSY appears 
 at the degenerate dyon point.
Therefore, this point is the unique and promising candidate 
 for the well-defined vacuum. 
Unfortunately, we have no knowledge about the correct description 
 of the effective theory around the degenerate singular point, 
 since the condensations of two BPS states well overlap there. 


 %
\end{document}